\documentclass[aps,prb,showpacs,groupedaddress,superscriptaddress,twocolumn]{revtex4}

\usepackage{graphicx}
\usepackage{amsmath,amssymb}
\usepackage{dcolumn}
\usepackage{bm}

\begin{document}

\title{Spin-current diode with a ferromagnetic semiconductor}

\author{Qing-Feng Sun}
\email{sunqf@iphy.ac.cn}
\affiliation{International Center for Quantum Materials, School of Physics, Peking University, Beijing 100871, China}
\affiliation{Collaborative Innovation Center of Quantum Matter, Beijing 100871, China}
\author{X. C. Xie}
\affiliation{International Center for Quantum Materials, School of Physics, Peking University, Beijing 100871, China}
\affiliation{Collaborative Innovation Center of Quantum Matter, Beijing 100871, China}

\date{\today}

\begin{abstract}
Diode is a key device in electronics: the charge
current can flow through the device under a forward bias, while almost no current flows under a reverse bias.  Here we propose a corresponding device in
spintronics: the spin-current diode, in which the forward spin
current is large but the reversed one is negligible. We show that the
lead/ferromagnetic quantum dot/lead system and the
lead/ferromagnetic semiconductor/lead junction can work as
spin-current diodes. The spin-current diode, a low dissipation device, may have important
applications in spintronics, as the conventional charge-current diode does in
electronics.
\end{abstract}


\maketitle

A new subdiscipline of condensed matter physics, spintronics that
exploits electron spin to replace the role of electron charge in
electronic devices, is emerging rapidly and generating great
interests in the last decade.\cite{ref1,ref2,ref2a} Some interesting effects, such as the giant magnetoresistance phenomena,\cite{ref3,ref4}
(quantum) spin Hall effect,\cite{ref5,ref6,ref6a,ref7,ref8,ref9} and
persistent spin current\cite{ref10,ref11,ref12} have been discovered. They received a lot
attention and even generated great applications. Based on the spin
degree of freedom of electron, many electronic devices have been
proposed, such as the magnetoresistance device,\cite{ref1,ref2,ref3,ref4,ref13,ref14}
the Datta-Das transistor,\cite{ref15,ref16} spin valve,\cite{ref17,ref18} spin filter,\cite{ref19,ref20,ref21} spin-polarized current
generator, pure spin-current generator\cite{ref6a,ref22,ref23,ref24,ref25} and spin diode device.\cite{ref26,ref27,ref28,ref29,ref30} Many
above-mentioned devices (e.g. the Datta-Das transistor,\cite{ref16} spin filter,\cite{ref19,ref20}
spin-current generator,\cite{ref6a,ref22,ref23,ref24,ref25} etc,) have been successfully realized in
experiments. Some extensive practical applications have also been achieved.
For example, the magnetoresistance devices have been extensively applied
in various commercial electronic products nowadays and generated
huge commercial values.\cite{ref3,ref4}

On the side of the charge degree of freedom of an electron, a very
important electronic device is the diode: the charge current
can flow through it under a forward bias but the current is
tiny under a reverse bias. In the last century,
various (charge-current) diodes have been successfully made.\cite{ref31} The
diode is a key device in electronics and it has been
extensively used in almost all electronic products and
instruments.\cite{ref31} A natural question is whether there exists a counterpart to
the charge-current diode on the side of the spin degree of freedom.
It means that a device has such a characteristic: a large spin
current can flow through the device under a forward spin bias,
while under a reverse spin bias the spin current is very
small. We name this device the spin-current diode, which should
have great applications in the future high-operating-speed and
energy-saving spintronic products. It also merits a mention that
some previous works have proposed and investigated the spin diode,\cite{ref26,ref27,ref28,ref29,ref30}
e.g. in the system consisting of normal-metal lead/quantum dot
(QD)/ferromagnetic (FM) lead, both theoretically\cite{ref27,ref28} and experimentally.\cite{ref29,ref30}
The spin diode in these previous references\cite{ref26,ref27,ref28,ref29,ref30} are devices in which the charge current or its spin polarization are strongly asymmetric for the forward and reverse biases, utilizing electron spin degrees of freedom. Thus, they are still devices dealing with the charge current. Therefore, the spin-current diode proposed here is completely different from the spin diode studied before, and it has never been explored before.

\begin{figure}[!htb]
\includegraphics[height=5.8cm, width=8cm, angle=0]{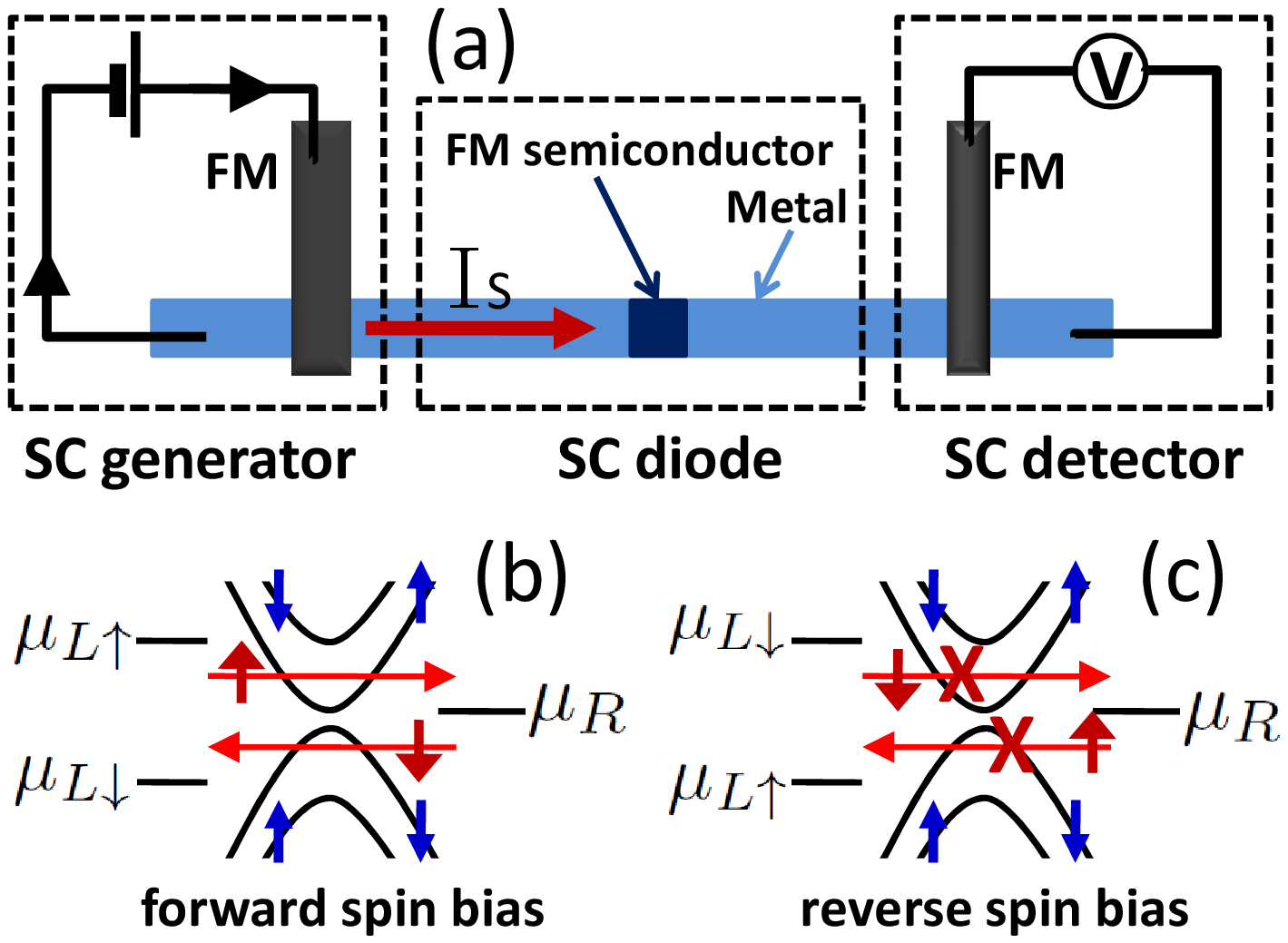}
\caption{(Color online) (a) The schematic diagrams for the spin-current generator,
the spin-current diode and the spin-current detector. Here the
device in the middle dotted box is the spin-current diode proposed
in this article. (b) [(c)] schematically shows that the spin current
can (can not) flow through the lead/FM semiconductor/lead junction
under the forward (reverse) spin bias.\label{fig1}}
\end{figure}

In this Letter, we propose a spin-current diode, which consists of a
normal lead/FM semiconductor (or FM QD)/normal lead.
Fig.~\ref{fig1}(a) shows schematically the device, consisting of the spin-current
generator, spin-current diode and spin-current detector, in three
dotted-line rectangles, respectively. The spin-current generator and
detector which consist of the FM electrodes weakly coupled to
normal-metal leads, have successfully been realized in recent
experiments.\cite{ref6a,ref23,ref24,ref25} Here we use them to generate the spin current (i.e.
as a spin bias) as well as to detect the spin current. In the middle rectangle in
Fig.~\ref{fig1}(a), it is the spin-current diode proposed in the
present letter, in which a large spin current flows through it under
a forward spin bias but the spin current is quite small for the
reverse spin bias, and the charge current is zero always. Let us first discuss why the device can work as
a spin-current diode. For a FM semiconductor with a FM magnetization
$M$, the conduction band and valence band will be spin-resolved as
shown in Fig.~\ref{fig1}(b) and (c). Here the highest valence band
and the lowest conduction band are usually spin opposite and there is a gap
between them. The FM semiconductor has been extensively studied in the
last three decades, and various FM semiconductors, such as GaMnAs
and InMnAs, have been found.\cite{ref32,ref33,ref34,ref34a} Considering a spin bias acts
on the lead/FM semiconductor/lead junction, the chemical potential
of the left lead is spin-resolved,\cite{ref35} due to the spin-current generator
on the left side as shown in Fig.~\ref{fig1}(a). For the forward
spin bias, the spin-up chemical potential $\mu_{L\uparrow}$ is
larger than the spin-down one $\mu_{L\downarrow}$ [see
Fig.~\ref{fig1}(b)]. It can drive the spin-up electrons flow
from the left lead through the FM semiconductor to the right lead and
the spin-down electrons flow in the reverse direction, leading a
large forward spin current. On the other hand, for the reverse spin bias, $\mu_{L\uparrow}<\mu_{L\downarrow}$ [see
Fig.~\ref{fig1}(c)], the spin-down (spin-up) electrons
can hardly tunnel from the left (right) lead to the middle FM
semiconductor.  This is because there is no spin-down (spin-up)
band in the spin-bias window, so that the spin current with the reverse spin bias is very
small. Furthermore, we will also show that the device of
lead/FM QD/lead can have the similar function as that of the
lead/FM semiconductor/lead, and both of them can work as the
spin-current diodes.

We first study the spin transport through the lead/FM QD/lead
device, whose Hamiltonian is:
\begin{eqnarray}
H &=& \sum\limits_{\alpha,k,\sigma} \epsilon_{\alpha k}
a^{\dagger}_{\alpha k\sigma} a_{\alpha k\sigma}
+\sum\limits_{\sigma} (\epsilon_d-\sigma M) d^{\dagger}_{\sigma}
d_{\sigma}
\nonumber\\
&& +\sum\limits_{\alpha, k, \sigma} \left(t_{\alpha}
a^{\dagger}_{\alpha k\sigma}d_{\sigma}+ H.c. \right),
\end{eqnarray}
where $a_{\alpha k\sigma }$ ($a^{\dagger}_{\alpha k\sigma }$) and $
d_{\sigma}$ ($d^{\dagger}_{\sigma }$) are the annihilation
(creation) operators of electron with spin $\sigma$
($\sigma=\uparrow ,\downarrow)$ in the lead $\alpha$ ($\alpha=L, R)$
and the QD, respectively. The QD has a single energy level
$\epsilon_d$ with the spin splitting energy $M$ due to the FM
magnetization or the Zeeman effect of an external magnetic field.
By using the non-equilibrium Green's function method, the particle
currents from the left lead through the QD to the right lead are:\cite{ref36,ref37}
\begin{equation}
I_{\sigma} = (1/\hbar)\int(d\epsilon/2\pi) T_{\sigma}(\epsilon)
\left(f_{L\sigma}-f_{R\sigma} \right).
\end{equation}
Here the transmission coefficient $T_{\sigma}(\epsilon) =\Gamma_L
\Gamma_R/[\epsilon-\epsilon_d+\sigma M)^2+\Gamma^2/4]$ with the
linewidth function $\Gamma=\Gamma_L+\Gamma_R$,
$\Gamma_{\alpha}=2\pi\rho_{\alpha} |t_{\alpha}|^2$ with $\rho_{\alpha}$ being the density
of state of the lead $\alpha$. The Fermi-Dirac
distribution
$f_{\alpha\sigma}=1/\{\exp[(\epsilon-\mu_{\alpha\sigma})/k_B T]+1\}$
at temperature $T$. After obtaining the particle current, the
spin current $I_s = (\hbar/2)(I_{\uparrow}-I_{\downarrow})$ and the
(charge) current  $I_e =e (I_{\uparrow}+I_{\downarrow})$ can be
calculated straightforwardly. Considering the action of the spin
bias $V_s$, the chemical potentials of the left lead
$\mu_{L\uparrow}=-\mu_{L\downarrow}=eV_s$ is spin-resolved,\cite{ref35}
but that of the right lead $\mu_{R\uparrow}=\mu_{R\downarrow}=eV_R$ is
spin-independent due to the spin-current generator in the left side
as shown in Fig.~\ref{fig1}(a). In the calculation, by considering
the open circuit, the charge current $I_e$ exactly is zero, from which the
voltage $V_R$ of the right lead can be obtained.

\begin{figure}[!htb]
\includegraphics[height=6.4cm, width=8cm, angle=0]{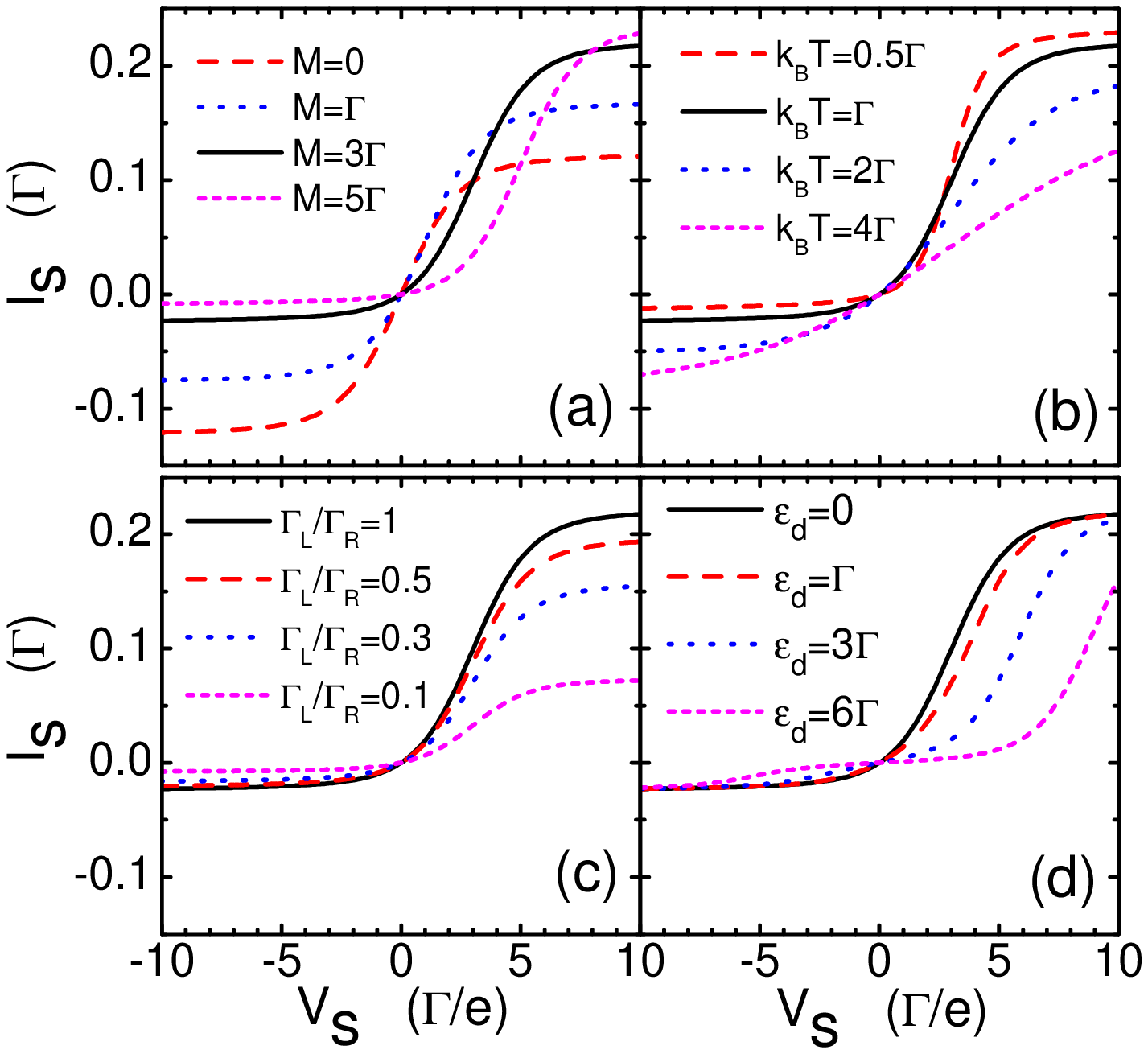}
\caption{(Color online) (a)-(d) show $I_s$ vs.
$V_s$ at the different $M$ (a), the different
temperature $k_B T$ (b), the different asymmetric couplings
$\Gamma_L/\Gamma_R$ (c) and the different QD's level $\epsilon_d$
(d). If the parameters are not shown in the legends, they are
$\epsilon_d=0$, $k_B T=\Gamma$, $M=3\Gamma$ and
$\Gamma_L/\Gamma_R=1$.\label{fig2}}
\end{figure}

Fig.~\ref{fig2} shows the spin current $I_s$ as a function of the
spin bias $V_s$. While the spin splitting energy $M=0$, $I_s(V_s)=
-I_s(-V_s)$ regardless of the other parameters. In this case, the device
does not have the diode features. On the other hand, while
$M\not=0$, the device shows the diode features, and the value of
spin current $I_s$ for the forward spin bias $V_s$ is not equal to
that of the negative $V_s$, i.e. $I_s(V_s)\not= -I_s(-V_s)$  [see
Fig.~~\ref{fig2}(a)]. For the larger spin splitting energy $M$, the
diode feature is more evident. For example, for $M=5\Gamma$, the
spin current can exceed $0.22\Gamma$ for the forward spin bias
$V_s=10\Gamma/e$, but it is only about $-0.008\Gamma$ for the reverse
spin bias $V_s=-10\Gamma/e$, and the ratio of the forward and
reverse spin current is about 30. In other words, the lead/FM QD/lead
device is a good spin-current diode. While $M$ is negative,
$I_s(-M,V_s) = -I_s(M, -V_s)$. This means that the polarity
direction of the spin-current diode can be reserved by reserving
$M$. For a positive $M$, the spin current can flow from the left
side through the device to its right side and the reversed spin current can
hardly flow through the device. However, for a negative $M$,
the reverse spin current can flow through the device and the
forward one can not. Compared to the charge-current diode, the ability to control the
polarity direction of the spin-current is a
great advantage. Notice that for the charge-current diode (e.g. p-n
junction), its polarity direction is fixed and can not be changed once the device is fabricated.

Let us discuss how the system parameters will affect the capability of
a spin-current diode. With increase of temperature $T$,
the spin current for a forward spin bias reduces but the reverse
one increases, and the diode feature gradually decays as shown in
Fig.~\ref{fig2}(b). For $k_B T < M$ (e.g. $k_B T=0.5\Gamma$ or
$\Gamma$), however, the diode feature is well maintained. 

So far we have considered cases that the coupling between QD and the left
and right leads are symmetric with $\Gamma_L =\Gamma_R$. In an
experiment, the couplings are usually asymmetric. Fig.~\ref{fig2}(c) shows
the effect of the asymmetric couplings on the diode feature of the
device. When the couplings are asymmetric, both the forward and
reverse spin currents fall, because the transmission
coefficients reduce. However, the ratio $|I_s(V_s)/ I_s(-V_s)|$ is
almost kept even for very large asymmetrical couplings, e.g.
$\Gamma_L /\Gamma_R=0.1$. This means that the device's diode feature
is robust against the asymmetrical couplings.

Thus far, we have set the QD's level $\epsilon_d$ at zero, {\i.e.} right at the Fermi
energy. But experimentally, the QD's level $\epsilon_d$ is
controllable by the gate voltage. Now, when $\epsilon_d$ is away from zero, we find that
the forward spin current at large forward spin bias and the
reverse spin current at large reverse spin bias can almost maintain
their values [see Fig.~\ref{fig2}(d)], thus, the diode feature holds well.
But the threshold spin bias $V_s^{th}$ increases with the
increase of $\epsilon_d$, and $V_s^{th}$ is roughly equal to
$\epsilon_d$. The forward spin current is small for $V_s<V_s^{th}$
and it quickly rises only when $V_s$ is beyond $V_s^{th}$. Another importance
message from Fig.~\ref{fig2}(d) is that the spin current can be sensitively enhanced by
adjusting the gate voltage, thus, the set-up provides the main function of a spin current transistor, in which an input charge signal at the gate voltage terminal can be changed to control the signal of the spin current.\\

Next, we study the spin transport through the lead/FM
semiconductor/lead junction. Here we consider that the FM
semiconductor has a conduction band and a valence band with the
energy gap $E_g$ and the FM magnetization $M$. Its Hamiltonian
is:
\begin{eqnarray}
 H_{SC}&=& \left(\begin{array}{ll}
 \frac{p_x^2 +p_y^2}{2\gamma_e m_0} -\hat{\sigma}_z M + \frac{E_g}{2} & 0\\
 0& - \frac{p_x^2 +p_y^2}{2\gamma_h m_0} -\hat{\sigma}_z M - \frac{E_g}{2}
 \end{array}\right) \nonumber \\
 &&+V(y),
\end{eqnarray}
where $p_{x/y} =-i\hbar \frac{\partial}{\partial x/y}$ is the
momentum operator, $\gamma_e m_0$ and $\gamma_h m_0$ are the
effective masses of the carriers in the conduction and valence bands
with the electron mass being $m_0$, and $\hat{\sigma}_z$ is a Pauli matrix.
The potential $V(y)=\infty$ when $y<0$ or $y>W$, and
$V(y)=0$ when $0<y<W$, namely, a FM semiconductor nanoribbon with
width $W$ is being considered. The energy bands of this FM semiconductor
are schematically shown in Fig.~\ref{fig1}(b) and (c).

First, let us assume that the transport from the FM semiconductor to
the lead is ballistic without the back-scattering. This assume is
reasonable if the lead has much more transverse modes than that in
the FM semiconductor.\cite{ref37} In this case, the transmission coefficients
$T_{\sigma}(\epsilon)$ are equal to the number of the transverse
modes of the FM semiconductor. After obtaining the transmission
coefficients, the spin current can be calculated as described
above. In the numerical calculations, we take the ribbon width
$W=20nm$, the gap $E_g = 0.5 eV$ which is a typical value for a
semiconductor, and $\gamma_e=\gamma_h=1$ (i.e. the conduction and
valence bands are mirror symmetrical with regard to the Fermi energy).

\begin{figure}[!htb]
\includegraphics[height=6.4cm, width=8cm, angle=0]{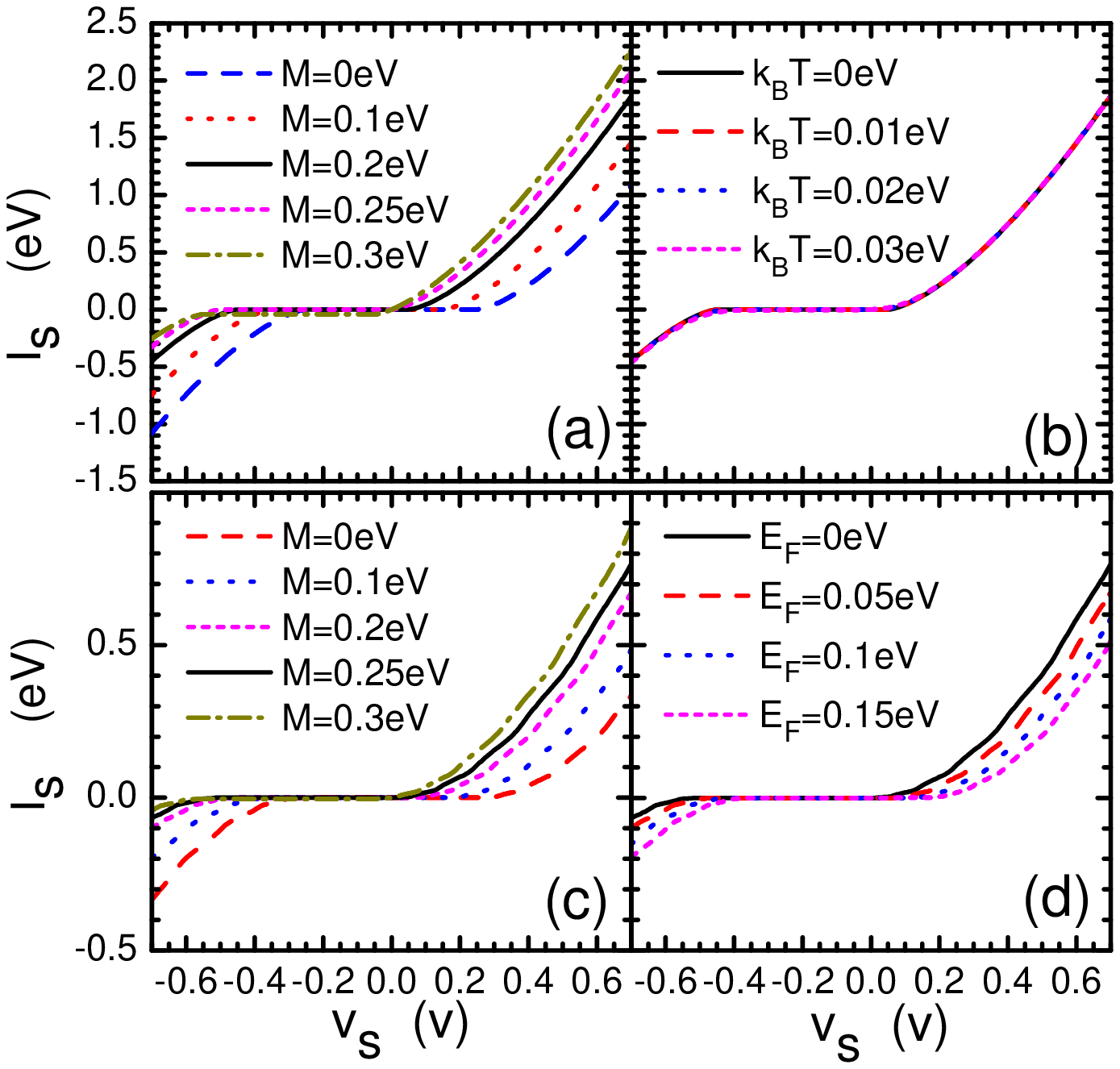}
\caption{(Color online) (a)-(d) show $I_s$ vs.
$V_s$ without [(a) and (b)] and with [(c) and (d)] the interface
scattering between the leads and the FM semiconductor. The FM
semiconductor's gap is $E_g=0.5eV$ and its size is with width
$W=20nm$ and length $5nm$. $k_B T=0$ for (a), (c)
and (d), $E_F=0$ for (a), (b) and (c), and $M=0.2eV$ (b) and $M=0.25eV$ (d). All curves in (b)
are almost overlap together.\label{fig3}}
\end{figure}

Fig.~\ref{fig3}(a) shows the spin current $I_s$ through the lead/FM
semiconductor/lead junction as a function of spin bias $V_s$ with
different FM magnetization $M$. While $M=0$, $I_s(V_s) =
-I_s(-V_s)$, and the rectification
of the spin current is absent from the device. When the magnetization $M$ is non-zero, the
forward spin current rises under the forward spin bias, but under
the reverse spin bias the reverse spin current is restrained, such that the
spin-current diode feature appears. Even for a small $M$, the
diode feature can be very pronounced. The reasons have been
described above in detail.

Let us discuss the characteristics of this spin-current diode. (1)
In a suitable spin bias range, the spin-current rectification of
this device is almost perfect. For the spin bias in the range of
$\frac{E_g}{2}-M <V_s <\frac{E_g}{2}+M$, the forward spin current
is in the order of $eV$. However, for the corresponding reverse spin
bias ($-\frac{E_g}{2}+M >V_s >-\frac{E_g}{2}-M$), the spin current
is almost zero. In this case, the ratio of the forward and reverse
spin currents, $|I_s(V_s)/I_s(-V_s)|$, is approaching infinity in principle
and the device works as an ideal spin-current diode. (2) The
threshold spin bias of the diode is $V_s^{th}=\max(\frac{E_g}{2}-M,
0)$, and the forward spin current appears until $V_s$ is larger than
$V_s^{th}$. (3) If the magnetization $M$ is larger than $E_g/2$, the
gap of the FM semiconductor is closed. In this case, the threshold
spin bias is zero.  Meanwhile, a small reverse spin current
appears. The diode feature is still quite good [see the curve of
$M=0.3eV$ in Fig.~\ref{fig3}(a)]. (4) As in the lead/FM QD/lead system,
the spin current in the lead/FM semiconductor/lead junction also has the
property, $I_s(-M,V_s) = -I_s(M, -V_s)$. So the polarity
direction of the spin-current diode can be reserved by tuning the
magnetization direction. Compared to the charge-current diode,
this is a great advantage. (5)
Fig.~\ref{fig3}(b) shows the spin current $I_s$ versus the spin bias
with different temperature $T$. Due to the presence of the gap,
the spin currents are almost independent of temperature. This means that the
diode feature of the device can be well kept even if $k_B T$ reaches to
$0.03eV$.

Second, we investigate what happens when the scattering exists at the
interface of the lead and the FM semiconductor, as is the case in a practical set-up.
In this case, the system can be described by the
Hamiltonian $H=H_{SC}+H_L+H_R+H_{C}$, where $H_{SC}$, $H_{L/R}$, and
$H_{C}$ are the Hamiltonians of the FM semiconductor, the left/right
lead, and the coupling between them, respectively. Here we consider
the non-magnetic normal leads with their Hamiltonians $H_{L/R}
=\frac{p_x^2+p_y^2}{2m_0}+ V_0$ and the potential energy $V_0$. The
Hamiltonian $H_{SC}$ has been shown in Eq.(3). In order to calculate
the transmission coefficients $T_{\sigma}$, we discretize the
Hamiltonian:\cite{ref37}
\begin{eqnarray}
&&H_{SC} = \sum\limits_{\bf i \in SC } a_{\bf i}^{\dagger} H_{{\bf i}{\bf i}}^{SC} a_{\bf i}
  +\sum\limits_{<{\bf i} {\bf j}> } a_{\bf i}^{\dagger} H_{{\bf i}{\bf j}}^{SC} a_{{\bf j}} ,\nonumber\\
&&H_{L/R} = \sum\limits_{\bf i \in L/R} b_{\bf i}^{\dagger} H_{{\bf i}{\bf i}}^{L/R} b_{\bf i}
  +\sum\limits_{<{\bf i} {\bf j}> } b_{\bf i}^{\dagger} H_{{\bf i}{\bf j}}^{L/R} b_{{\bf j}} ,\nonumber\\
&&H_{C} = \sum\limits_{i_y} ( b^{\dagger}_{0i_y} H_{LC}a_{1i_y}
 + b^{\dagger}_{N+1,i_y} H_{RC}a_{N i_y} +h.c.
 ), \nonumber
\end{eqnarray}
where $a_{\bf i}^{\dagger }$ ($a_{\bf i}$) and $b_{\bf i}^{\dagger
}$ ($b_{\bf i}$) are the creation (annihilation) operators in the FM
semiconductor and the leads, with the discrete site ${\bf i}
=(i_x,i_y)$. $i_x\in(-\infty,0)$, $i_x\in(1,N)$ and
$i_x\in(N+1,\infty)$ are the region of the left lead, the FM
semiconductor and the right lead, respectively. $H_{{\bf i}{\bf
i}}^{SC}=diag(\frac{E_g}{2}-M +4t_e, \frac{E_g}{2}+M +4t_e,
-\frac{E_g}{2}-M -4t_h, -\frac{E_g}{2}+M -4t_h)$ and $H_{{\bf i}{\bf
i}}^{L/R}=diag(4t_{L/R} +V_0, 4t_{L/R} +V_0)$ are the diagonal
matrices. $\langle {\bf i}{\bf j} \rangle$ stands for the nearest-neighbor sites
${\bf i}$ and ${\bf j}$, and $H_{{\bf i}{\bf j}}^{SC}=diag(-t_e,
-t_e, t_h, t_h)$ and $H_{{\bf i}{\bf j}}^{L/R}=diag(-t_{L/R},
-t_{L/R})$ describe the nearest neighbor hopping Hamiltonians. Here
the hopping elements $t_{e/h}=\frac{\hbar^2}{2\gamma_{e/h} m_0 a^2}$
and $t_L=t_R=\frac{\hbar}{2m_0 a^2}$ with the discretized lattice
constant $a$. The coupling Hamiltonians  $H_{LC}=H_{RC}=\left(\begin{array}{llll} t_c &
0&t_c &0\\ 0& t_c& 0&t_c\end{array}\right)$. After discretizing the
Hamiltonian, by using the non-equilibrium Green's function method,
the transmission coefficients $T_{\sigma} =Tr({\bf \Gamma}_L {\bf
G}^r {\bf \Gamma_R} {\bf G}^a)$ can be obtained straightforwardly.\cite{ref37,ref38}
In the following calculations, we take the FM semiconductor's width
$W=20nm$ and length $5nm$, and the lattice constant $a=0.1nm$. The gap is
set at $E_g = 0.5 eV$, $\gamma_e=\gamma_h=1$ and $t_L=t_R=t_e=t_h=t_c$.

Fig.~\ref{fig3}(c) shows the spin current $I_s$ versus the spin bias $V_s$ at
different FM magnetization $M$ in the presence of the interface
scattering. Due to the interface scattering, $I_s$ is
slightly smaller than the ideal case. However, the spin-current
diode feature holds well with a non-zero $M$: a large
spin current appears in the forward spin bias but the reverse spin current
is very small with the reverse spin bias. Furthermore, the five
aforementioned characteristics are all maintained in the present case. At last, we
consider the case of non-zero Fermi energy $E_F$ (i.e. it is away from the
particle-hole symmetric position). The Fermi energy can be tuned by
the gate voltage in the experiment. While $E_F\not=0$,
$\mu_{L\uparrow}=eV_s +E_F$, $\mu_{L\downarrow}=-eV_s +E_F$ and
$\mu_{R}=eV_R +E_F$. The results show that the threshold spin bias
$V_s^{th}$ slightly increases when $E_F$ deviates from zero. But the
spin-current diode features can survive when $E_F <E_g/2$, for
example, the values of the forward spin currents are evidently
larger than that of the reverse one for $E_F=0.1eV$ and $0.15eV$
[see Fig.~\ref{fig3}(d)]. Moreover, Fig.~\ref{fig3}(d) shows that
the spin current depends sensitively on the Fermi energy
or gate voltage, a main function of a spin current transistor.\\

In conclusion, we predict that the lead/FM semiconductor (FM QD)/lead
junction can work as the spin-current
diodes, through which the forward spin current can easily flow while
the reverse one is effectively blocked. In addition, the device
discussed above can be easily fabricated, thus, the proposed
spin-current diode is realizable with today's technology. The
spin-current diode may have important applications in spintronics,
just as the (charge-current) diode did in electronics.\\

{\textbf{Acknowledgments:}} This work was financially
supported by NBRP of China (2012CB921303 and 2012CB821402) and
NSF-China under Grants Nos. 11274364 and 91221302.\\

\end{document}